\documentclass[twocolumn,aps,prd,epsf]{revtex4}
\usepackage{epsfig,color}

\begin{document}

\title{The anti-decuplet candidate $\Xi^{--}$(1862)
as a heptaquark with the overlap of two anti-kaons and a nucleon}
\author{P. Bicudo}
\email{bicudo@ist.utl.pt}
\affiliation{Dep. F\'{\i}sica and CFIF, Instituto Superior T\'ecnico,
Av. Rovisco Pais, 1049-001 Lisboa, Portugal}
\begin{abstract}
I study the very recently discovered S=-2 resonance $\Xi^{--}$ (1886) by the NA49 
collaboration at the CERN SPS. This resonance was already predicted,
with a mass close to 1.9 GeV, in a recent publication mostly dedicated to the 
S=1 resonance $\Theta^+$(1540). To confirm this recent prediction, I apply the 
same standard quark model with a quark-antiquark annihilation constrained by chiral 
symmetry. 
This method also explained with success the repulsive hard core of nucleon-nucleon, 
kaon-nucleon exotic scattering, and the short range attraction present in pion-nucleon
and pion-pion non-exotic scattering. I find that 
repulsion excludes the $\Xi^{--}$ as a $dd ss \bar u$ s-wave pentaquark. I
explore the $\Xi^{--}$ as a heptaquark, equivalent to a $N+K+K$
linear molecule, with positive parity and total isospin
$I=3/2$. I find that the kaon-kaon repulsion is cancelled by the
attraction existing in the kaon-nucleon channels. In our framework this
state is easier to bind than the $\Theta^+$ described by a $\pi+K+N$
borromean bound-state. The remaining $I=1/2$ doublet and $I=1$ triplet
of the exotic anti-decuplet are also studied, and the coupling to p-wave decay
channels is briefly addressed.
\end{abstract}
\maketitle

\section{introduction}

\par
In this paper I study the exotic hadron $\Xi^{--}$
(narrow hadron resonance of 1862 MeV decaying into a $\Xi^- \pi^-$) very
recently discovered by the NA49 collaboration at the CERN SPS 
\cite{Alt,Fischer,Price}. 
The simultaneous discovery of a resonance with similar
mass and width decaying into a $\Xi^- \pi^+$ provides evidence
for a $I=3/2$ iso-quadruplet. 
This completes a missing link of the anti-decuplet
\cite{Chemtob,Praszalowicz,Diakonov1}
which includes the recently discovered $\Theta^{+}$(1540)
\cite{Nakano,Barmin,Stepanyan,Barth,Hyodo}. 
Moreover pentaquark structures have been observed in the lattice
both with parity + and with parity -
\cite{Csikor,Sasaki,Chiu}.

\par
The $\Theta^+$ is an extremely exciting state,
because it may be the first exotic hadron to be discovered, with quantum 
numbers that cannot be interpreted as a quark and an anti-quark meson or 
as a three quark baryon. 
In what concerns the quark structure it was known for a long time that
the simplest and lightest s-wave multiquarks would be repulsive and very
unstable. 
For instance we recently computed
\cite{Bicudo00} 
the mass of the groundstate s-wave $I=0$, $J^P=1/2^+$ $uudd \bar s$ pentaquark, 
and we checked that it would have a mass of 1535 MeV, close to $M_{\Theta^+}$.  
However in this channel we find a purely repulsive exotic $N-K$ 
hard core s-wave interaction
\cite{Bender,Bicudo,Barnes}. 
This suggests why pentaquarks have been hard to find, 
and at the same time this indicates that pentaquarks should include an excitation.
Because excited multiquark systems are indeed difficult to study, different 
perspectives of the pentaquarks are welcome to fully address these new states.

\par 
Exotic multiquarks are expected since the early works of Jaffe
\cite{Jaffe,Strottman, Sorba,Roisnel}. 
Soon after the $\Theta^+$ was observed,
Jaffe and Wilczek
\cite{Jaffe2}, 
and Karliner and Lipkin
\cite{Karliner} 
proposed that the pentaquarks are arranged in microscopic coloured diquarks or 
triquarks, connected by a string in a p-wave state. This is
a very appealing structure, in particular the p-wave system
tends to have a narrow width because the decay is only possible if the diquarks
overlap. This model needs a novel cancellation of some of the mass of diquarks, 
to compensate the large p-wave excitations.

\par
The exotic anti-decuplet was first predicted, with the correct $\Theta^+$ mass 
and similar decay width, by Diakonov, Petrov and Polyakov
\cite{Diakonov1}. 
These authors interpret the exotic anti-decuplet as a 
rotational excitation of the chiral topological soliton
\cite{Chemtob,Praszalowicz,Diakonov1}. 
This approach suggests that chiral symmetry and p-wave excitations are crucial
to understand the pentaquarks. However the Skyrme Chiral Soliton has some 
difficulty to reproduce the short range repulsion in N-N interactions. 
 
\par
Soon after the $\Theta^*$ was discovered, we proposed another pentaquark model 
\cite{Bicudo00} 
which estimates a mass of 1.9 Gev for the S= -2, Q = -2 state. Importantly,
this predicted mass is quite close to the observed one. To motivate our 
proposal let me first review the five possible excitations of the quark model, 
in decreasing energy shift order. 
The first excitation is the radial one, with an
energy shift of $M_{\rho^*(1--)}-M_{\rho(1--)}\simeq 700$MeV.
The second excitation is the angular one, with an
energy shift of $M_{f_1^{(1++)}}-M_{\rho^{(1--)}}\simeq 500$MeV.
The third excitation is the spin one, with an
energy shift of $M_{K^{*(1--)}}-M_{K^{(0-+)}}\simeq 400$MeV.
The fourth excitation is the flavour one, with an
energy shift of $M_{K^{*(1--)}}-M_{\rho^{(1--)}}\simeq 150$MeV
or $M_{\omega^{(1--)}}-M_{\rho^{(1--)}}\simeq 10$MeV.
The fifth excitation is the quark-antiquark pair creation one, with an
energy shift of $M_{\pi^{(0-+)}} \simeq 140$MeV.
The light mass $M_{\theta^+}\simeq M_N+M_K + 100$MeV of the $\theta^+$ pentaquark 
suggests that it either has a flavour excitation or a $q- \bar q$ pair creation,
and not a p-wave excitation. 
Because the production processes show no evidence for a weaker flavour changing 
reaction, we explored 
\cite{Bicudo00}
the $q- \bar q$ creation hypothesis.
Moreover, when a flavor singlet quark-antiquark pair $u \bar u +d \bar d$ 
is created in the pentaquark $H$, the resulting crypto-heptaquark 
$H'$ is a state with an opposite 
parity to the original $H$, where the reversed parity occurs due to the intrinsic 
parity of fermions and anti-fermions. In this sense the new heptaquark $H'$ can be 
regarded as the chiral partner of $H$. And, because $H'$ is expected to have 
the lowest possible mass, it is naturally rearranged in a 
baryon belonging to the s-wave baryon octet and in two pseudoscalar mesons belonging 
to the s-wave meson octet. In this approach the mass of the heptaquark $H'$ is simply 
expected to be slightly lower than the exact sum of these standard hadron masses due 
to the negative binding energy. 
For instance we recently suggested
\cite{Bicudo00} 
that the $\Theta^+$ is probably a 
$K-\pi-N$ molecule with binding energy of 30 MeV,
a borromean three body s-wave boundstate of a $\pi$, a $N$ and a $K$
\cite{Bicudo00,Llanes-Estrada,Kishimoto}, with
positive parity 
\cite{Oh}
and total isospin $I=0$.
We also addressed the S= -2, Q = -2 state $\Xi^{--}$, 
suggesting that it is a $\bar K - N- \bar K$ molecule. The 
NA49 result for the iso-quadruplet of 1.862 GeV is consistent with 
a binding energy of 60 MeV for the hadronic molecule.

\par
In this paper I extend the quantitative techniques used in our first 
publication for the $\Theta^+$,
\cite{Bicudo00}
to the remaining of the anti-decuplet. 
I start by reviewing, in Section \ref{framework}, the standard Quark Model (QM), 
and the Resonating Group Method (RGM)
\cite{Wheeler,Ribeiro}
which is adequate to study states where several quarks overlap. 
Using the RGM, I show that the corresponding exotic baryon-meson short 
range s-wave interaction is repulsive in exotic channels and 
attractive in the channels with quark-antiquark annihilation. 
In most iso-multiplets, except in 
the $S=-1$ iso-multiplet the short range repulsion contradicts
a possible pentaquark with a narrow width.
Section \ref{binding} proceeds with the study of the heptaquarks 
in the exotic anti-decuplet,
which are bound by the attractive non-exotic $\bar K-N$ and $\bar K-K$
interactions. I compute the masses of the exotic anti-decuplet. 
In Section \ref{coupling} the coupling and decays to p-wave channels are
addressed. Finally the conclusion is presented in Section \ref{conclusion}. 

\section{framework}
\label{framework}

\par
Our Hamiltonian is the standard QM Hamiltonian,
\begin{equation}
H= \sum_i T_i + \sum_{i<j} V_{ij} +\sum_{i \bar j} A_{i \bar j} \,
\label{Hamiltonian}
\end{equation}
where each quark or antiquark has a kinetic energy $T_i$ with a
constituent quark mass. The colour dependent two-body
interaction $V_{ij}$ includes the standard QM confining term and a
hyperfine term. The quark-antiquark annihilation-creation potential
$A_{i \bar j}$ is necessary when the potential complies with chiral
symmetry, including the light pion mass and the Adler Zero
\cite{Bicudo0,Bicudo1,Bicudo3,Bicudo4}.
The RGM provides an accurate framework
\cite{Wheeler,Ribeiro}
to compute the effective multiquark energy 
using the matrix elements of the quark-quark interactions. 
Any multiquark state can be decomposed in antisimmetrised combinations 
of simpler colour singlets, the baryons and mesons. 
For the purpose of this paper the details of the potentials in eq. 
(\ref{Hamiltonian}) are unimportant, only their matrix elements, 
extracted from the baryon spectroscopy, matter
\begin{eqnarray}
\langle V_{hyp} \rangle &\simeq& \frac{4}{3} \left( M_\Delta-M_N \right) \ ,
\nonumber \\
\langle A \rangle_{S=0} &\simeq&- {2 \over 3} \left( 2M_N-M_\Delta \right) \ .
\end{eqnarray}

\par
The detailed calculations are similar to the ones in reference
\cite{Bicudo00}, 
and lead to the attraction/repulsion criterion,
\\
- {\em whenever the two interacting hadrons have quarks (or antiquarks)
with a common flavour, the repulsion is increased by the Pauli principle,
\\
- when the two interacting hadrons have a quark and an
antiquark with the same flavour, the attraction 
is enhanced by the quark-antiquark annihilation}.
\\
In the particular case of one nucleon interacting with anti-kaons 
and with kaons, this implies that the short range $K-N$ and $\bar K -\bar K$ 
interactions are repulsive, while the short range $\bar K -K$ and $\bar K -N$
interactions are attractive. Quantitatively
\cite{Bicudo00,Bicudo2,Bicudo5},
the effective potentials computed for the channels relevant to this
paper, are
\begin{eqnarray}
 & V_{K-N}=& 
{ {1 \over 4} + {1 \over 6}  ( {\vec \tau}_K + {\vec \tau}_N)^2 
\over 1 - {1 \over 6}  ( {\vec \tau}_K + {\vec \tau}_N)^2 }
\langle V_{hyp} \rangle  |\phi_{\bf 0}^\alpha><\phi_{\bf 0}^\alpha|
\ , 
\nonumber \\
& V_{K-\bar K}=& { 2 +\sigma -( {\vec \tau}_K + {\vec \tau}_{\bar K})^2  
\over 6} \langle A \rangle \,  |\phi_{\bf 0}^\alpha><\phi_{\bf 0}^\alpha|
\ , 
\nonumber \\
 & V_{\bar K-N}=& { 3 -( {\vec \tau}_{\bar K} + {\vec \tau}_N)^2  
 \over 6} \langle A \rangle \, |\phi_{\bf 0}^\alpha><\phi_{\bf 0}^\alpha|
\ , 
\nonumber \\
 & V_{\bar K-\bar K}=&
{ {1 \over 4}( {\vec \tau}_{\bar K} + {\vec \tau}_{\bar K})^2     
\over 1- {1 \over 12}( {\vec \tau}_{\bar K} + {\vec \tau}_{\bar K})^2    }
 \langle V_{hyp} \rangle  |\phi_{\bf 0}^\alpha><\phi_{\bf 0}^\alpha|
\ , 
\label{overlap kernel}
\end{eqnarray}
where ${\vec \tau}$ are the isospin matrices, normalised with ${\vec \tau \,}^2= \tau(\tau+1)$. 
In the chiral limit one would expect that the $I=0\, , V_{\bar K-N}$ cancels with the
$I=1, \ V_{\bar K - \bar K}$, and this is confirmed by eq. (\ref{overlap kernel}).
To arrive at eq.\ref{overlap kernel} we used a harmonic oscillator basis $|\phi_{\bf n}^\alpha>$
for the multiquark wave-function ,
where the inverse hadronic radius $\alpha$ is the only free parameter in this framework.

\section{binding in the anti-decuplets}
\label{binding}

\par 
The simplest pentaquarks are not 
expected to bind due to the attraction/repulsion criterion. For instance
the $\Xi^{--}$ cannot be a $ddss\bar u$ pentaquark. The possible elementary
color singlets $(dds)-(s\bar u)$ or  $(dss)-(d\bar u)$ are repelled because
the elementary color singlets share the same flavour $d$ or $s$. This 
also implies that the $\pi-\Xi$ and $\bar K-\Sigma$ systems are unbound.
Then the only way to have attraction consists in adding at least one 
quark-antiquark pair to the system. 

\par
However including an extra pion in the fundamental configurations is not
possible, except in the $\Theta^+$. 
Because the pion is very light it is not expected to bind into a narrow
resonance
\cite{Rupp1}, 
except in the $\Theta^+$ where it may be attracted both by a $K$ 
and a $N$ in a borromean structure. Moreover the $K$ and $N$ are repelled
in this framework, and the narrow $\Theta^+$ is not supposed to exist unless it
includes a pion to bind it. 

\par
Nevertheless, in the other iso-multiplets, where a $\bar K$ exits, binding 
is possible because the $\bar K$ is attracted both by the $K$ 
and the $N$. Therefore, although the $\bar K - \bar K$ system is repulsive,
the $K-\bar K-N$, $\bar K-N$ and $\bar K-N-\bar K$ systems are expected to
bind. It is then convenient to build the anti-decuplet like a combinatoric 
Newton pyramid, starting by the summit, $uudd\bar s$, I=0, $\Theta^+$.
To reach any of the other three iso-multiplets in the anti-decuplet,
one simply needs to add respectively one, two or three 
$I=1/2$, $\bar K$,  ( $s\bar u$,  $s\bar d$ ) to the $\Theta^+$.
Altough we advocate 
\cite{Bicudo00}
that $\Theta^+$ is a $K-\pi-N$ linear molecule, 
let us consider for the flavour purpose, that it has the 
flavour of a I=0, $K-N$ system. 

\par 
The next $I=1/2$, $N^*$  iso-multiplet ( $uud$ , $udd$ ) can
be obtained combining a $I=1/2$, $\bar K$,  ( $s\bar u$,  $s\bar d$ )
with the $\Theta^+$. A possible binding structure is a 
$K-\bar K-N$ linear molecule. 

\par 
The next $I=1$, $\Sigma^*$  iso-multiplet ( $uus$ , $uds$ , $dds$ ) can
be obtained combining a $I=1/2$, $\bar K$,  ( $s\bar u$,  $s\bar d$ )
with the $N^*$. The simplest exotic binding structure is pentaquark with
the flavour of a $\bar K-N$ system. 

\par
Finally the last $I=3/2$, $\Xi^*$  iso-multiplet 
( $uuss\bar d$, $uss$, $dss$, $ddss\bar u$ ) 
can be obtained combining a $I=1/2$, $\bar K$,  ( $s\bar u$,  $s\bar d$ )
with the $\Sigma^*$. In this case the simplest exotic binding structure 
is a $\bar K-N-\bar K$ linear molecule. The different multiquarks
are summarised in Table \ref{heptaquark candidates}.

\subsection{The I=1 pentaquark}

I now compute the energy of the simplest state, the I=1 iso-triplet $\Sigma^*$, as
a pentaquark with the quantum numbers of a $\bar K-N$ system. 
The other $\bar K -N$ system, the iso-singlet $I=0$ $\Lambda(1405)$ has been
studied in detail in the literature. Nevertheless the $I=1$ system
is also attractive, moreover the $\bar K-N$ binding is relevant to the $I=3/2$, $\Xi^*$.
In this case the reduced mass is $\mu=325$ MeV, and the potential 
$ v | \phi_{000}^\alpha \rangle \langle  \phi_{000}^\alpha| $
is attractive and separable, where 
$v=-71.5 MeV$ was computed in eq. (\ref{overlap kernel}) 
and where $\alpha$ is a free parameter. 
In the $\bar K - N$ case, binding exists if $\alpha <\sqrt{-4 \mu \, v}=$ 304 MeV. 
However for a binding energy of the order of 60 MeV, close to the
one proposed for the the exotic $\Xi^{--}$ and consitent with a
crypto-exotic $\Sigma^*$(1385) a too small $\alpha = 74 MeV $ 
would be required. Therefore binding is expected in this system, although an 
accurate prediction of the binding energy of this $\Sigma^*$ system would need 
technical improvements of our method, see Section \ref{conclusion} for details.

\subsection{The I=1/2 and I=3/2 heptaquarks}

%
%
\begin{table}[t]
\begin{tabular}{c|c|c}
channel                    & flavour $m_I$          & multiquark   \\
\hline
$I=0 \ , S=1             $ & $uudd\bar s$           & $K-\pi-N$ molecule \\
$ I={1 \over 2} \ , S=0  $ & $uud\, , udd  $        & $K-\bar K-N$ molecule  \\
$ I=1 \ , S=-1           $ &  $uus\, , uds\, , dds$ & $\bar K-N$  pentaquark \\
$ I={ 3 \over 2} \ , S=-2  $ &  $uuss\bar d\, , uss\, , dss\, , ddss\bar u $ 
& $\bar K-N-\bar K$ molecule \\
\hline
\end{tabular}
\caption{ Proposed list of multiquark states, candidates to the exotic pentaquark and
heptaquark anti-decuplets.}
\label{heptaquark candidates}
\end{table}

\par
I now use an adiabatic Hartree method to study the stability of the
linear $I= 3/2 \ \ \bar K -N-\bar K$ molecule. While the $I=1$ pentaquark
is not a molecule since the five quarks and antiquarks overlap in a 
s-wave state, here the large $\bar K -\bar K$ repulsion, with $v=234$ MeV, 
prevents the overlap of all the quarks. Essentially the wave-function
of the $N$ is centred between the two $\bar K$, and the two $\bar K$
only overlap with the nucleon, but not with each other. This results in
a linear molecule. I solve a Schr\"odinger
equation for a $\bar K$ in the potential produced by a nucleon placed
at the origin and by the other $\bar K$ placed at a distance $-{ \bf a}$ 
of the nucleon. 
The potential of the $N$ is produced by a $\bar K$ anti-kaon at the point
$-{ \bf a}$ and another anti-kaon at the point ${+\bf a}$. The 
potentials are respectively,
\begin{eqnarray}
V_{\bar K}&=& v_{\bar K-N} |\phi_{\bf 0}^\alpha><\phi_{\bf 0}^\alpha|
+ v_{\bar K-\bar K}|\phi_{-{\bf a}}^\alpha><\phi_{-{\bf a}}^\alpha| \ ,
\label{potential K}
\\
V_N&=& v_{\bar K-N} |\phi_{-{\bf a}}^\alpha><\phi_{-{\bf a}}^\alpha|
+ v_{\bar K-N} |\phi_{{\bf a}}^\alpha><\phi_{{\bf a}}^\alpha| \ ,
\label{potential N}
\end{eqnarray} 
where the sub-index denotes the position of the potential.
This produces three binding energies $E_{\bar K}, \, E_{\bar K}, \, E_N$,
and three wave-functions. In the Hartree method the total energy is the 
sum of these energies minus the matrix elements of the potential energies,
\begin{equation}
E=2 E_{\bar K} + E_N - 2 <\phi_{\bar K}|V_{\bar K}|\phi_{\bar K}>
-  <\phi_N|V_N|<\phi_N> .
\label{Hartree energy}
\end{equation}
This is easily computed once the two Schr\"odinger equations are respectively solved
with the potentials (\ref{potential K}) and (\ref{potential N}).
The total energy is a function of the distance $\bf a$, and I minimise it as 
a function of $| {\bf a} |$. The energy minimum is obtained for 
$ | {\bf a} | \simeq 0.7 / \alpha$. The centres
of the two $\bar K$ are separated by a distance of $ 1.4/\alpha$, and therefore
they essentially do not overlap. I first verify that binding is easy to produce in this 
system, although an $\alpha$ consistent with the expected nucleon radius $\leq 0.7$ Fm
would produce a small binding energy.
Then the case of the excessively small $\alpha = 74 MeV $, that provides a large binding 
for the $\bar K-N$ system, is investigated. 
This would already correspond to an extremely large nucleon mean radius $\alpha^{-1}$,
nevertheless let us check that a larger binding can be obtained for the $\bar K -N-\bar K$.
In this case I find that all the three one hadron binding energies are similar to 
-40 MeV. When the potential energies are subtracted in eq. (\ref{Hartree energy}), 
the final binding energy results in $E\simeq$ -40MeV which is consistent 
with a $\Xi^{--}$ mass of $1,88$ GeV. This remains larger than a mass of $1,86$ GeV 
for the $\Xi^{--}$, however a complete computation of the effect of coupled channels 
on the binding energy remain to be estimated. The length $1.4/\alpha$ of the 
$\bar K -N-\bar K$ linear molecule is larger than $1$ Fm, even when we have
a small nucleon, with $\alpha=304$ MeV. 

\par
A similar binding energy, and a similar mass of the order of $1.9$ GeV can be 
computed for the $K-\bar K- N$ system in the non-exotic iso-doublet.

\section{Coupling and decays to p-wave channels}
\label{coupling}

\par
In the case of  the $\Xi^{--}(1860)$, the mass is already
400 MeV higher than the mass of the $\Xi^{-}+ \pi^{-}$ system,
and 200 MeV higher than the mass of the $\Sigma^{-}+ K^{-}$ 
system. In this case the p-wave excitation may also be relevant,
as suggested by the Chiral Soliton model and by the Diquark model. 
Moreover the decay channels are also p-wave channels. This
motivates the study of the coupling of the s-wave 
crypto-heptaquark system to the p-wave pentaquark systems. 

\par
\begin{figure}[t]
\epsfig{file=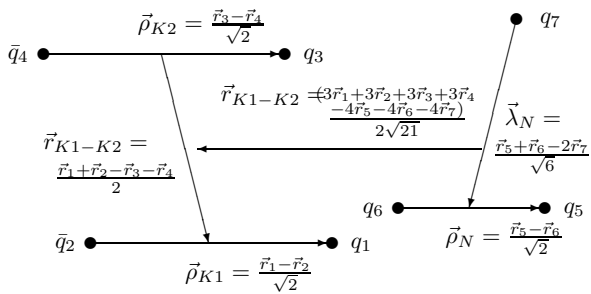,width=12cm}
\caption{
The Jacobi coordinates of the heptaquark.
}
\label{Jacobi coordinates}
\end{figure}

The simplest model to couple a s-wave crypto-heptaquark to a
p-wave pentaquark consists in assuming the standard $^3P_0$ 
quark-antiquark creation/annihilation potential. The coupling
form factor is computed with the overlap of the s-wave 
crypto-heptaquark $\bar K + N + \bar K$ with the p-wave pentaquarks 
$\bar K + \Sigma$ (or $\pi + \Xi$) plus a flavour singlet $^3P_0$ 
quark-antiquark pair. Between these two wave-functions we must 
sandwich the quark antisimmetriser. After separating the center of 
mass coordinates, the coupling is equivalent to an overlap of harmonic
oscillator wave-functions of the six Jacobi coordinates depicted in 
Fig. \ref{Jacobi coordinates}.
%
%
The Ribeiro graphical
\cite{Ribeiro3}
rules, and the harmonic oscillator energy 
conservation, require that at least one of the heptaquark coordinates 
should include a radial excitation. The $\bar K - \bar K$ relative 
coordinate is expected to be partly excited because the  $\bar K - \bar K$ 
are repelled. If we compare our coupling with the decay 
of the rho into two pions we find a suppression in the coupling by a 
factor of 1/3. The decay width is proportional to the square of the coupling 
and this factor already explains the small decay width of heptaquarks 
which are found in all experiments to be smaller than 20 MeV. 
Moreover the Ribeiro rules show that overlaps with excited wave-functions are 
further suppressed. However, for a consistent computation, the probability 
for a $^3P_0$ annihilation to occur in a pentaquark remains to be consistenly
studied. Nevertheless the p-wave pentaquark components should essentially not 
affect the mass of the crypto-heptaquark.

\par
The relative comparison of the $\Theta^+$ and $\Xi^{--}$ decay widths
can be estimated with a better precision than the full computation of the decay. 
The decay widths depend on the phase space, and they are proportional 
to $ \mu \sqrt{ 2 \mu E}$ in the non-relativistic case and to $E^2$ in the
ultra-relativistic case, when the momentum is smaller 
that the inverse radius $\alpha$. 
Therefore I expect that the partial decay width of the $\Xi^{--}$ to a 
p-wave $\bar K - \Sigma$ is larger than the  $\Theta^+$ decay to a 
p-wave $K - N$ by a factor of $1.5$, and I expect that the partial 
decay width of the $\Xi^{--}$ to a p-wave $\pi - \Xi$ is larger
than the  $\Theta^+$ decay to a p-wave $K - N$ by a factor smaller than $2$.
Thus it is expected that the total decay width of the $\Xi^{--}$ is 
larger by a factor $\simeq 3$ than the decay width of the $\Theta^+$. 
Nevertheless, because the decay width of the $\Theta^+$ is quite small, the
coupling of the crypto-heptaquark to the p-wave pentaquark is not expected to
be large. 

%
%
\begin{figure}[t]
\epsfig{file=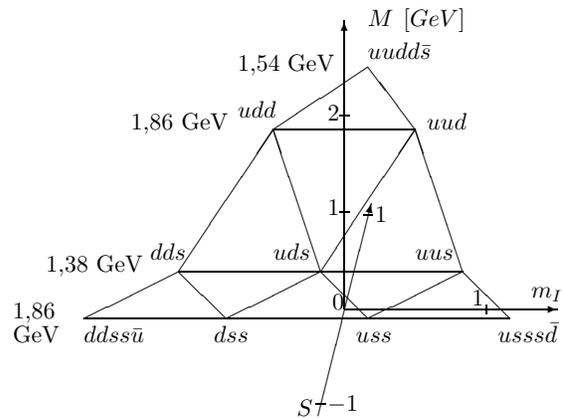,width=12cm}
\caption{
The masses of the narrow exotic anti-decuplets, in the heptaquark and 
pentaquark scenarios, are shown in a three dimensional 
strangeness-flavour-mass plot.
The $\Theta^+$ iso-singlet is a $K-\pi-N$ molecule. 
The iso-doublet $N^*$ is a $K-\bar K-N$ molecule. 
The iso-triplet $\Sigma^*$ is a $\bar K-N$ pentaquark. 
The $\Xi_5^*$ iso-quadruplet is a $\bar K-N-\bar K$ molecule. 
}
\label{triangle}
\end{figure}
%
%
%

\section{Conclusion}
\label{conclusion}

\par
To conclude, in this paper I address the exotic anti-decuplets 
with a standard quark model Hamiltonian, where the quark-antiquark
annihilation is constrained by the spontaneous breaking of chiral
symmetry. I first derive a criterion showing that the 
$\Theta^+$ and $\Xi^{--}$ hadrons very recently discovered cannot 
be an s-wave or p-wave pentaquark. It is plausible
that the $\Xi^{--}$ is a linear $\bar K - N - \bar K $ molecule,
a heptaquark state. The $I=0$, $\Theta^+$ and the $I=1/2$, $N^*$
are also heptaquarks, or linear meson-meson-baryon molecules.
In these linear molecules the two external hadrons do not overlap.
All these heptaquarks have a positive parity.
The only pentaquark is the iso-triplet $\Sigma^*$, with the quantum numbers
of a tightly bound s-wave $\bar K - N$ system, a negative parity system.
The spectrum of the pentaquark and heptaquark antidecuplets is depicted 
in Figure \ref{triangle}, assuming that the experimental $M_\Theta=1.54$ GeV 
and $M_\Xi=1.86$ GeV are correct. Importantly, the $\Xi^{--}$(1862) is expected 
to decay both to the $\bar K - \Sigma$ and to the $ \pi - \Xi$ p-wave channels.

\par 
For a future improvement of this work it is important to discuss the size 
parameter $\alpha^{-1}$. In our previous paper
\cite{Bicudo00}
we used a quite large $\alpha\simeq 10$ Fm$^{-1}$, for the potential involving 
the pion. 
The pion is expected to couple with a shorter range interaction than other 
hadrons, because the Adler zero suppresses the low momentum part of the 
pionic couplings. Nevertheless the very short range pion potential and the longer 
range kaon potential suggest that the hadron-hadron interactions should have at 
least two different interactions. 
For instance the N-N interaction is decomposed in a short range interaction 
due to quark Pauli repulsion, a medium range $\sigma$ or Two Pion Exchange 
Potential, and a long range One Pion Exchange Potential. The use of 
similar interactions may further increase the binding of multiquark molecules.
Other technical improvements that deserve to be investigated,
although they are not straightforward to implement at the microscopic level
of quarks, are the coupling to p-wave channels and the exact three body equations.

\par
I expect that the advocated heptaquarks will help to motivate the detailed
experimental studies of the $\Theta_5$, $N^*$, $\Sigma^*$ and $ \Xi^*_5$
\cite{Huang,Liu,Nakayama}.
This will be crucial to discriminate the different theoretical models candidate 
to explain the anti-decuplet,
\cite{Bicudo00,Jaffe2,Karliner,Diakonov1,Close,Cheung,Cohen,Jennings,Matheus,Carlson,Bijker,Ellis,Pakvasa}.

\acknowledgements
I thank Gon\c{c}alo Marques for discussions on the algebraic computations of 
this paper. As a note added in proof, I would like to acknowledge Takashi 
Nakano and Eulogio Oset for mentioning that a $\Sigma^*$ similar to the one 
advocated here has already been predicted
\cite{Jido,Sarkar,Kolomeitsev, Sarkar2},
and that Cristoph Hanhart pointed that a $\pi-\pi-N$ heptaquark may correspond
to the Roper resonance $N^*$(1440) 
\cite{Krehl}. This supports the heptaquark $\bar K - N - \bar K$ model for the
$\Xi^{--}$(1862).



\begin{thebibliography}{00}
%
\bibitem{Alt}
C.~Alt {\it et al.}  [NA49 Collaboration],
arXiv:hep-ex/0310014.
%
\bibitem{Fischer}
H.~G.~Fischer and S.~Wenig,
arXiv:hep-ex/0401014.
%
\bibitem{Price}
J.~W.~Price, J.~Ducote, J.~Goetz and B.~M.~K.~Nefkens  [CLAS Collaboration],
arXiv:nucl-ex/0402006.
%
\bibitem{Chemtob} M.~Chemtob,
Nucl.\ Phys. {\bf 256}, 600 (1985).
%
\bibitem{Praszalowicz} M.~Praszalowicz,
in \textit{Skyrmions and Anomalies}, M.~Jezabek and M.~Praszalowicz, eds.,
World Scientific (1987), 112. 
%
\bibitem{Diakonov1}
D.~Diakonov, V.~Petrov and M.~V.~Polyakov,
Z.\ Phys.\ A {\bf 359} 305 (1997) [arXiv:hep-ph/9703373].
%
\bibitem{Nakano}
T.~Nakano {\it et al.}  [LEPS Collaboration],
Phys.\ Rev.\ Lett.\  {\bf 91}, 012002 (2003)
[arXiv:hep-ex/0301020].
%
\bibitem{Barmin}
V.~V.~Barmin {\it et al.}  [DIANA Collaboration],
Phys.\ Atom.\ Nucl.\  {\bf 66}, 1715 (2003)
[Yad.\ Fiz.\  {\bf 66}, 1763 (2003)]
[arXiv:hep-ex/0304040].
%
\bibitem{Stepanyan}
S.~Stepanyan {\it et al.}  [CLAS Collaboration],
Phys.\ Rev.\ Lett.\  {\bf 91}, 252001 (2003)
[arXiv:hep-ex/0307018].
%
\bibitem{Barth}
J.~Barth {\it et al.}  [SAPHIR Collaboration],
arXiv:hep-ex/0307083.
%
\bibitem{Hyodo}
T.~Hyodo, A.~Hosaka and E.~Oset,
arXiv:nucl-th/0307105.
%
\bibitem{Csikor}
F.~Csikor, Z.~Fodor, S.~D.~Katz and T.~G.~Kovacs,
JHEP {\bf 0311} (2003) 070
[arXiv:hep-lat/0309090].
%
\bibitem{Sasaki}
S.~Sasaki,
arXiv:hep-lat/0310014.
%
\bibitem{Chiu}
T.~W.~Chiu and T.~H.~Hsieh,
arXiv:hep-ph/0403020.
%
\bibitem{Bicudo00}
P.~Bicudo and G.~M.~Marques,
arXiv:hep-ph/0308073.
%
\bibitem{Bender}
I.~Bender, H.~G.~Dosch, H.~J.~Pirner and H.~G.~Kruse,
Nucl.\ Phys.\ A {\bf 414}, 359 (1984). 
%
\bibitem{Bicudo}
P.~Bicudo and J.~E.~Ribeiro,
Z.\ Phys.\ C {\bf 38}, 453 (1988); 
P.~Bicudo, J.~E.~Ribeiro and J.~Rodrigues,
Phys.\ Rev.\ C {\bf 52}, 2144 (1995). 
%
\bibitem{Barnes}
T.~Barnes and E.~Swanson, Phys.\ Rev.\ C {\bf 49} 1166 (1994);
J.~S.~Hyslop, R.~A.~Arndt, L.~D.~Roper and R.~L.~Workman,
Phys.\ Rev.\ D {\bf 46} 961 (1992).
%
\bibitem{Jaffe}
R.L.~Jaffe, SLAC-PUB-1774, talk presented at the
 Topical Conf. on Baryon Resonances, Oxford, England, July 5-9, 1976;
R.~L.~Jaffe,
Phys.\ Rev.\ D {\bf 15} 281 (1977).
%
\bibitem{Sorba} 
H.~Hogaarsen and P.~Sorba,
Nucl.\ Phys. {\bf B145}, 119 (1978).
%
\bibitem{Strottman}
D.~Strottman,
Phys.\ Rev.\ D {\bf 20} 748 (1979).
%
\bibitem{Roisnel} 
C.~Roiesnel, Phys.\ Rev. {\bf D20}, 1646 (1979).
%
\bibitem{Jaffe2}
R.~L.~Jaffe and F.~Wilczek,
Phys.\ Rev.\ Lett.\  {\bf 91}, 232003 (2003)
[arXiv:hep-ph/0307341].
%
\bibitem{Karliner} 
M.~Karliner and H.~J.~Lipkin,
Phys.\ Lett.\ B {\bf 575}, 249 (2003)
[arXiv:hep-ph/0402260],
M.~Karliner and H.~J.~Lipkin,
arXiv:hep-ph/0307343.
%
\bibitem{Llanes-Estrada}
F.~J.~Llanes-Estrada, E.~Oset and V.~Mateu,
arXiv:nucl-th/0311020.
%
\bibitem{Kishimoto}
T.~Kishimoto and T.~Sato,
arXiv:hep-ex/0312003.
%
\bibitem{Oh}
Y.~s.~Oh, H.~c.~Kim and S.~H.~Lee,
Phys.\ Rev.\ D {\bf 69}, 014009 (2004)
[arXiv:hep-ph/0310019];
Y.~s.~Oh, H.~c.~Kim and S.~H.~Lee,
arXiv:hep-ph/0312229.
%
\bibitem{Wheeler}
J.~Wheeler, Phys.\ Rev.\ {\bf 52}, 1083 (1937); {\bf 52}, 1107
(1937).
%
\bibitem{Ribeiro}
J.~E.~Ribeiro, Z.\ Phys.\ C {\bf 5}, 27 (1980).
%
\bibitem{Bicudo0}
P.~Bicudo, 
Phys.\ Rev.\ C {\bf 67}, 035201 (2003). 
%
\bibitem{Bicudo1}
P.~Bicudo, S.~Cotanch, F.~Llanes-Estrada, P.~Maris, J.~E.~Ribeiro and
A.~Szczepaniak,
Phys.\ Rev.\ D {\bf 65}, 076008 (2002) [arXiv:hep-ph/0112015].
%
\bibitem{Bicudo3}
P.~Bicudo and J.~E.~Ribeiro,
Phys.\ Rev.\ D {\bf 42}, 1611 (1990); 
1625 (1990); 
1635 (1990). 
%
\bibitem{Bicudo4}
P.~Bicudo,
Phys.\ Rev.\ C {\bf 60}, 035209 (1999). 
%
\bibitem{Bicudo2}
P.~Bicudo, M.~Faria, G.~M.~Marques and J.~E.~Ribeiro,
Nucl.\ Phys.\ A {\bf 735}, 138 (2004)
[arXiv:nucl-th/0106071].
%
\bibitem{Bicudo5}
P.~Bicudo,
arXiv:hep-ph/0401106.
%
\bibitem{Rupp1}
E. van Beveren, C. Dullemond, C. Metzger, J. E. Ribeiro, T.A. Rijken, and 
G. Rupp,
Z. Phys. C 30, 615 (1986).
%
\bibitem{Ribeiro3}
J.~E.~Ribeiro, Phys. Rev. D {\bf 25}, 2406 (1982);
E.~van~Beveren, Zeit. Phys. C {\bf 17}, 135 (1982);
J.~E.~Ribeiro,
arXiv:hep-ph/0312035.
%
\bibitem{Huang}
P.~Z.~Huang, W.~Z.~Deng, X.~L.~Chen and S.~L.~Zhu,
arXiv:hep-ph/0311108.
%
\bibitem{Liu}
Y.~R.~Liu, P.~Z.~Huang, W.~Z.~Deng, X.~L.~Chen and S.~L.~Zhu,
arXiv:hep-ph/0312074.
%
\bibitem{Nakayama}
K.~Nakayama and K.~Tsushima,
Phys.\ Lett.\ B {\bf 583}, 269 (2004)
[arXiv:hep-ph/0311112].
%
\bibitem{Close}
F.~ Close  {\em in} the closing talk of
Hadron2003, Aschaffenburg, Germany (2003).
%
\bibitem{Cheung} 
K.~Cheung,
arXiv:hep-ph/0308176.
%
\bibitem{Matheus}
R.~D.~Matheus, F.~S.~Navarra, M.~Nielsen, R.~Rodrigues da Silva and S.~H.~Lee,
Phys.\ Lett.\ B {\bf 578}, 323 (2004)
[arXiv:hep-ph/0309001].
%
\bibitem{Jennings}
B.~K.~Jennings and K.~Maltman,
arXiv:hep-ph/0308286.
%
\bibitem{Cohen}
T.~D.~Cohen and R.~F.~Lebed,
Phys.\ Lett.\ B {\bf 578}, 150 (2004)
[arXiv:hep-ph/0309150].
%
\bibitem{Carlson}
C.~E.~Carlson, C.~D.~Carone, H.~J.~Kwee and V.~Nazaryan,
Phys.\ Lett.\ B {\bf 579}, 52 (2004)
[arXiv:hep-ph/0310038];
C.~E.~Carlson, C.~D.~Carone, H.~J.~Kwee and V.~Nazaryan,
arXiv:hep-ph/0312325.
%
\bibitem{Bijker}
R.~Bijker, M.~M.~Giannini and E.~Santopinto,
arXiv:hep-ph/0310281.
%
\bibitem{Ellis}
J.~Ellis, M.~Karliner and M.~Praszalowicz,
arXiv:hep-ph/0401127.
%
\bibitem{Pakvasa}
S.~Pakvasa and M.~Suzuki,
arXiv:hep-ph/0402079.
%
\bibitem{Jido}
D. Jido, J.A. Oller, E. Oset, A. Ramos, U.G. Meissner, 
Nucl.\  Phys.\  A {\bf 725} , 181 (2003). 
[arXiv:NUCL-TH 0303062]
%
\bibitem{Sarkar}
S.~Sarkar, E.~Oset and M.~J.~Vicente Vacas,
arXiv:nucl-th/0407025.
%
\bibitem{Kolomeitsev}
E.E. Kolomeitsev, M.F.M. Lutz, 
Phys.\ Lett.\  B {\bf 585} , 243 (2004). 
arXiv:NUCL-TH 0305101.
%
\bibitem{Sarkar2}
Sourav Sarkar, E. Oset, M.J. Vicente Vacas, 
arXiv:NUCL-TH 0404023.
%
\bibitem{Krehl}
O.~Krehl, C.~Hanhart, S.~Krewald and J.~Speth,
Phys.\ Rev.\ C {\bf 62}, 025207 (2000)
[arXiv:nucl-th/9911080].
%
\end{thebibliography}
\end{document}